\documentclass[aps,prl,twocolumn,floatfix, groupedaddress]{revtex4}
\usepackage{graphicx}

\begin{document}

\title{The thermodynamic spin magnetization of strongly correlated 2D electrons in a silicon inversion layer}
\author{O.\ Prus$^1$, Y.\ Yaish$^1$, M.\ Reznikov$^1$, U.\ Sivan$^1$, and V.\ Pudalov$^2$ }
\affiliation{$^1$Dep. of Physics and Solid State Institute,
Technion-IIT, Haifa 32000, Israel}

\affiliation{$^2$P.\ N.\ Lebedev Physics Institute, 119991 Moscow,
Russia}
\begin{abstract}
A novel method, invented to measure the minute thermodynamic
magnetization of dilute two dimensional fermions, is applied to
electrons in a silicon inversion layer. The interplay between the
ferromagnetic interaction and disorder enhances the low
temperature susceptibility up to 7.5 folds compared with the Pauli
susceptibility of non-interacting electrons. The magnetization
peaks in the vicinity of the density, where transition to strong
localization takes place. At the same density, the susceptibility
approaches the free spins value (Curie susceptibility), indicating
an almost perfect compensation of the kinetic energy toll
associated with spin polarization by the energy gained from the
Coulomb correlation. Yet, the balance favors a paramagnetic phase
over spontaneous magnetization in the whole density range.
\end{abstract}

\maketitle

\section{Introduction}
The nature of the ground state of degenerate two dimensional (2D)
fermions at zero magnetic field is an outstanding open problem,
which has not been deciphered despite decades of research. In the
absence of disorder the ground state is believed to be determined
by an interplay between the kinetic energy, $ E_{F}$, and the
inter-particle interaction energy, $E_{c}=e^{2}/\kappa a$, where
$a=(\pi n)^{-1/2}$ is the inter particle distance, $n$ is the
areal particle density, and $\kappa $ is the host dielectric
constant. The relative importance of the two energy scales is
characterized by $r_{s}=a/a_{0}$, with $ a_{0}$ being the Bohr
radius. For electrons in a single band $r_{s}=E_{c}/E_{F}$, while
for the (100) surface of silicon $r_{s}=E_{c}/2E_{F}$ due to the
two-fold valley degeneracy. At very high densities ($r_{s}\ll 1$)
a 2D system approaches the non-interacting degenerate gas
paramagnetic limit, characterized by the Pauli susceptibility
$\chi _{0}$. As the density is reduced, the growing ferromagnetic
correlations lead to substantial enhancement of the spin
susceptibility $\chi $. The system is predicted to remain
paramagnetic up to $r_{s}\approx 20\div 25$, where numerical
calculations~\cite{Senatore,attaccalite} find a quantum phase
transition to a ferromagnetic liquid phase~\cite{Don}. At lower
density, $ r_{s}\approx 37$~\cite{T&C}, the Coulomb correlations
are predicted to lead via another phase transition to a quantum
Wigner crystal with frustrated antiferromagnetic spin
arrangement~\cite{Ceperley_2001} followed by transition to a
ferromagnetic arrangement at an even lower
density~\cite{Voelker01}. The energy balance between the ferro and
paramagnetic states is very subtle and the density window where
ferromagnetism may take place is small
~\cite{Senatore,attaccalite}. Such a ferromagnetic phase has never
been observed experimentally. The situation is fundamentally
complicated by the unavoidable disorder present in any realistic
system. In the absence of Coulomb interactions all wave functions
of a 2D system are believed to be exponentially localized~\cite
{gang4}. Localization modifies the Coulomb interaction
dramatically in the low density limit. The interplay between
kinetic energy, interaction, and disorder was worked out
theoretically for the case of relatively weak
disorder~\cite{aa,Fin,Zala}. It was found that the interaction
suppresses the localizing effect of disorder, especially in the
presence of valley degeneracy~\cite{Punnoose}. Yet, at low enough
densities disorder prevails and localization always commences.

Notwithstanding the substantial research done thus far, there is
presently no agreed picture of the phase diagram corresponding to
a realistic 2D fermion system. It is clear that the spin degree of
freedom plays a crucial role in the low density regime, $n\leq
2\times 10^{11}~{\rm cm^{-2}}$, but the minute total magnetic
moment pertaining to such a small number of spins has hindered,
thus far, any direct measurement of the thermodynamic spin
magnetization. Present estimates of the 2DEG magnetization in
silicon rely on susceptibility data obtained from transport
measurements, either Shubnikov-de Haas (Sh-dH) oscillations in a
tilted magnetic field ~\cite{FS68,OH99,VZ1,gm} or saturation of
the magnetoresistance in an in-plane field~\cite{SKDK,VZ2}. The
two approaches led to contradicting conclusions. While the
magnetoresistance data were interpreted as indicating the long
awaited Bloch-Stoner~\cite{Don} instability at the critical
density for the metal-insulator transition, analysis of Sh-dH
oscillations points against such instability~\cite{Pudalov:02}.

At the heart of the present manuscript is a novel method invented
to measure the thermodynamic magnetization directly. We apply the
method to a high mobility 2D electron layer in silicon. In
particular, we find that, as the density is reduced, the weak
field spin susceptibility is progressively enhanced up to
7.5~$\chi _{0}$, but the ferromagnetic instability is never
realized. The system turns insulating before it polarizes and
electron localization then leads to a reduction in the Coulomb
interaction. The localization transition is thus characterized by
a sharp cusp in magnetization. Interestingly, we find indications
for localized magnetic moments in coexistence with the itinerant
electrons, even at high carrier densities.

\section{Method, samples, and experimental setup}
\begin{figure}
\includegraphics [width=8cm] {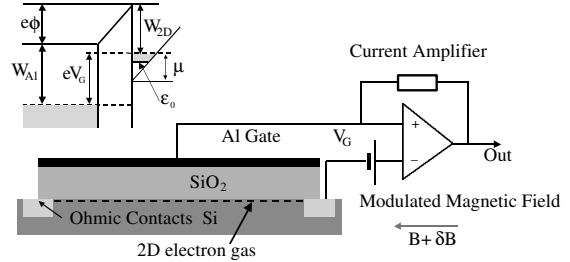}
\caption{Magnetization measurement setup and band diagram of the
2D confining potential} \label{setup}
\end{figure}

The experimental setup is presented in Fig.~\ref{setup}. An
external bias, $V_{G}$, sets a constant electrochemical potential
difference between the gate and the 2D channel equal to the sum of
the electrostatic potential difference, $\phi$, and the difference
between the aluminum gate and the 2DEG work-functions, $W_{Al}$
and $W_{2D}$, respectively
\begin{equation} \label{VG}
eV_{G}=e\phi(n)+W_{Al}-W_{2D}(n,B).
\end{equation}
Modulation of the in-plane magnetic field by an auxiliary coil at
a frequency $\omega $ modulates the chemical potential of the 2DEG
and, hence, $W_{2D}$ ($W_{Al}$ modulation is negligible). Since
$V_G$ is kept constant, the differential of Eq.~\ref{VG} vanishes.
The 2DEG chemical potential, $\mu$, equals the ${\rm Si-SiO_2}$
band discontinuity minus $W_{2D}$ (Fig.~\ref{setup}).
Consequently, one obtains
\begin{equation} \label{d}
e \frac{\partial{\phi}}{\partial{n}}dn
+\frac{\partial{\mu}}{\partial{n}}dn+
\frac{\partial{\mu}}{\partial{B}}dB=0,
\end{equation}

or
\begin{equation}\label{dmudb}
\frac{\partial{\mu}}{\partial{B}}=-\left(e\frac{\partial{\phi}}{\partial{n}}+
\frac{\partial{\mu}}{\partial{n}}\right)\frac{dn}{dB},
\end{equation}

\noindent where $(e \partial{\phi}/ \partial{n} +
\partial{\mu}/\partial{n})/e^2=C^{-1}$ is the independently measured inverse
capacitance per unit area, comprising the geometrical and chemical
potential contributions. The latter contribution includes well
width and interaction effects.

In terms of the induced current, $\delta I$, and the magnetic
field modulation, $\delta B$,  one obtains
\begin{equation}\label{dmudb1}
\frac{\partial{\mu}}{\partial{B}}=-\frac{ie\delta I} {C\omega
\delta B}.
\end{equation}

\noindent Since the 2D layer thickness and the screening length
are minuscule compared with the oxide thickness, the capacitance
is close to the geometrical one, and hence, constant to within 1\%
in the whole density range. Using one of Maxwell's relations,
$\partial M /\partial n=-\partial\mu/\partial B$, we obtain
$\partial M /\partial n$ and integrate it numerically with respect
to $n$ to derive the magnetization $M(B,n)$. The magnetic
susceptibility $\chi $ is calculated from the slope of $M(B,n)$
versus $B$ at small fields. An additional constant field, induced
by the main coil, facilitates magnetization measurements at finite
magnetic field.

While the method is conceptually straightforward, its realization
is demanding since the current induced by the field modulation is
typically on the order of $ 10^{-15}$~A, while the spurious
current induced in all wire loops by the ac magnetic field, and
even more so by mechanical vibrations of the sample in the dc
magnetic field, are potentially larger by several orders of
magnitude. The induced currents were minimized in the experiment
by careful compensation of all loops. Mechanical vibrations were
minimized by rebuilding the relevant parts of the refrigerator to
achieve high enough mechanical rigidity. After building several
prototypes we were able to drive all mechanical resonances to
frequencies considerably higher than the field modulation
frequency, and hence, eliminate the mechanical vibrations at the
measurement frequency.

The same setup can be used to measure the much larger orbital
magnetization in a perpendicular magnetic field. The method's
sensitivity scales with sample's area and magnetic field
modulation amplitude. For the $4~{\rm mm^2}$ sample and 0.03~T rms
field modulation used here, it was about 10$^{-14}$~J/T,
comparable to the best sensitivity achieved with SQUID-based
magnetometers~\cite{Meinel:APL97} (a review of the current state
of the art in magnetization measurements can be found
in~\onlinecite{Awshalom00}). The advantage of our
(and~\onlinecite{Meinel:APL97}) method is in its applicability to
arbitrary magnetic fields and temperatures, as well as to a wide
range of conductivities. The extraction of the magnetization from
Sh-dH oscillations, on the other hand, requires perpendicular
magnetic fields, low temperatures and high enough mobilities. The
most interesting regime, the transition to strong localization is,
hence, at the limit of its reach.

The samples used in the experiment were similar to those used in
Refs.~\onlinecite{rhoT1,rhoT2}. They consisted of 5\thinspace mm
long 0.8\thinspace mm wide Hall bars with 2.5\thinspace mm
separation between the potential probes. The oxide was
200\thinspace nm thick, leading to $C=678$~pF device capacitance.
We applied $-15$~V bias to the substrate in order to minimize the
contacts resistance~\cite{Prus02}. The maximal mobility under this
bias reached $17,000 \thinspace {\rm cm^{2}/Vs}$ . An alternating
magnetic field of typically 100-300~G~(rms) and $f=\omega /2\pi
=5\div 20$~Hz was applied parallel to the layer along with a
desired constant field. A preamplifier with $\sim 2~fA/\sqrt{Hz}$
current noise was used to measure the current and bias the gate
(Fig.~1).

Unlike Sh-dH based measurements, our method is sensitive to the
total thermodynamic magnetization comprising the spin part as well
as the diamagnetic orbital contribution due to the finite ($ \sim
50$~\AA) thickness of the 2D layer (Fig.~\ref{setup}). Localized
states also contribute to the measured magnetization, as long as
they exchange particles with the 2DEG at a rate faster than
$\omega$.

\section{Experimental results and discussion}
\begin{figure}[tpb]
\includegraphics [width=8cm] {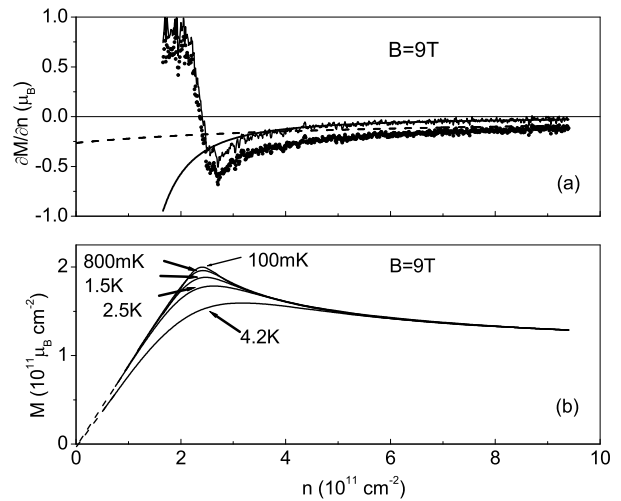}
\caption{(a) Total and spin $\partial M /\partial n$. Dots -
measured total $\partial M /\partial n$, dashed line - diamagnetic
contribution, thick smooth line - spin $\partial M /
\partial n$ extracted from Sh-dH data~\cite{gm}, thin solid line - spin $\partial M /
\partial n$ extracted by subtraction of the diamagnetic contribution from the total
$\partial M /\partial n$. (b) Spin magnetization obtained by
integration of $\partial M /\partial n$ data. The dashed line
demonstrates that the curves extrapolate to zero $M$ at vanishing
density, as they should.} \label{bardat}
\end{figure}

The measured $\partial M /\partial n$ at 9~T magnetic field and
$T=100$~mK is depicted by dots in Fig.~\ref{bardat}a. The smooth
solid line depicts the same quantity as extracted from Sh-dH
data~\cite{gm}. The difference between the two curves is
attributed to the diamagnetic shift due to the sub-band energy
level ($\varepsilon_0$ in Fig.~\ref{setup}) dependence upon the
magnetic field and to the presence of localized spins. Both
effects do not appear in Sh-dH oscillations. While the localized
spins are certainly relevant to the study of spin magnetization in
real samples, the diamagnetic part reflects an orbital effect,
which is outside the scope of our interest. To estimate the latter
contribution we assume that the Sh-dH measurements at high
densities (say $n\geq 5\times 10^{11}~{\rm cm^{-2}}$), where the
number of localized spins is small, give the spin magnetization
correctly. The diamagnetic contribution is then given by the
difference between our measured thermodynamic magnetization and
the one extracted from the Sh-dH data. At zero density, on the
other hand, one can calculate the diamagnetic shift in a
single-particle picture. To complete the estimate at intermediate
densities we interpolate between the two limits to obtain the
dashed line in Fig.~\ref{bardat}a. The spin magnetization in the
whole density range is obtained by subtraction of the diamagnetic
contribution (dashed line) from the measured data (dots). The
effect of the magnetic field, even at 9~T, on the subband energy
$\varepsilon_0$ is much smaller than the inter-subband spacing.
Therefore, the diamagnetic contribution to the magnetization
should depend linearly upon magnetic field, in accordance with our
high density data. The overall diamagnetic contribution in the
low-density range is small compared with the spin contribution.
Moreover, it varies slowly with density. The extracted spin
magnetization is therefore only slightly affected by the details
of the interpolation procedure. Yet, for $n\leq
2\times10^{11}~{\rm cm^{-2}}$ we find that the saturation value of
the extracted spin magnetization $\partial M /\partial n$ (solid
line in Fig.~\ref{bardat}a) is lower by $\approx10 $ \% than the
one Bohr magneton per electron, expected for full polarization.
Since all spins are likely to be polarized at low density and 9~T,
we attribute the discrepancy to an underestimate of the
diamagnetic contribution at low densities. The error in our
measured data is much smaller than 10 \%. The magnetization values
presented below may, hence, underestimate the actual magnetization
at low densities by up to $2 \times 10^9~\mu_B~{\rm cm^{-2}}$ per
Tesla. This uncertainty is immaterial for our conclusions. The
rest of the paper focuses on the spin contribution obtained by the
above procedure.

The spin magnetization (Fig.~\ref{bardat}b) at a given field is
obtained by numerical integration of the extracted $\partial M
/\partial n$ values with respect to $n$. Since the magnetization
can be measured only above a certain density, for which the sample
resistance is lower than $\simeq{\rm 1~M\Omega}$, the integration
cannot start from zero density, where $M=0$. Consequently, our
integration yields the magnetization up to a constant, which is
chosen so that the magnetization at high densities equals the
values extracted from the Sh-dH oscillations. We neglect the small
number of localized spins, which are present even at high
densities. The fact that the resulting curves at all temperatures
extrapolate to practically zero magnetization at $n=0$ (dashed
lines in Fig.~\ref{bardat}b) confirms that the integration
constants are chosen properly.

\begin{figure}[tpb]
\includegraphics [width=8cm] {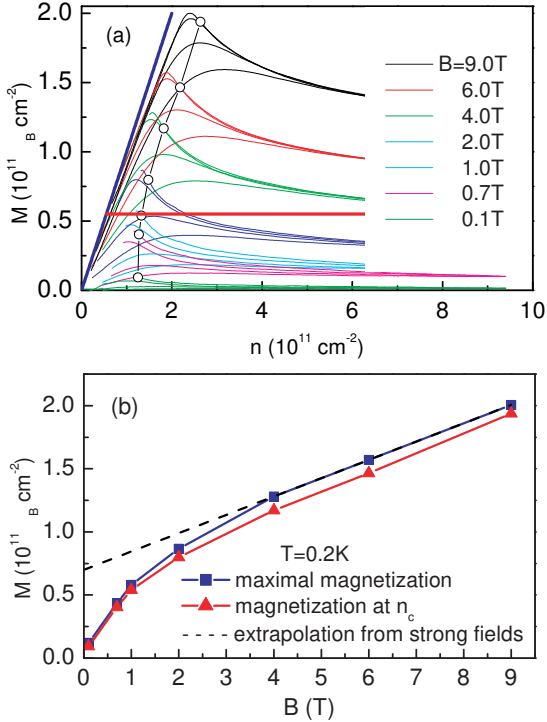}
\caption{(a) Spin magnetization as a function of density at
different magnetic fields and temperatures 0.2, 0.8, 2.5 and
4.2~K; higher magnetization corresponds to lower temperature.
Critical densities, $n_c$, are marked by circles. Thick blue line
- full magnetization, thick red line - magnetization of a
degenerate ideal electron gas at $B=6$~T. (b) Maximal spin
magnetization and spin magnetization at the critical densities
plotted against magnetic field. Dashed line - extrapolation from
high magnetic fields.} \label{M(n)}
\end{figure}

The spin magnetization at various magnetic fields and four
temperatures is depicted in Fig.~\ref{M(n)}a. For each magnetic
field the curves with higher magnetization values correspond to
lower temperatures. The thick blue line corresponds to full
polarization of all carriers at a given density and the thick red
line to the zero temperature magnetization of a non-interacting
degenerate electron gas at $B=6$~T. The empty circles denote for
each magnetic field the critical density $n_c(B)$, which separates
the metallic regime from the insulating one~\cite{AbrKravSar}. At
higher densities the resistance decreases as the temperature is
reduced (metallic behavior) while at lower densities it increases
(insulator). Whether the metallic behavior indicates a true 2D
metal or merely finite temperature transport through localized
states with long enough localization length is presently an open
question. It is clear, though, that the insulating regime
corresponds to localized states (either in the sense of
percolation or in the sense of exponentially decaying wave
functions) with progressively smaller localization lengths at
lower densities. At high magnetic fields, full spin alignment
persists up to densities considerably higher than those predicted
for non-interacting electrons (compare the non-interacting and the
experimental curves for $B=6$~T). Curiously, for all magnetic
fields the magnetization reaches its maximal value at densities
only slightly lower than the critical density $n_c$. As more
carriers are added to the layer, the total magnetization is
monotonically reduced. The large negative slope of the curves in
the vicinity of $n_c$ indicates that the added delocalized
electrons prefer to occupy the upper spin subband. At still higher
densities the magnetization is further reduced towards the
respective non-interacting values.

Fig.~\ref{M(n)}b depicts the maximal magnetization as well as the
magnetization at the critical densities versus magnetic field. The
data set an upper limit on the zero field polarization at the
critical density, $n_c\approx 1.25 \times 10^{11}~{\rm cm^{-2}}$,
to less than $2 \times 10^{10}~{\rm \mu_b~cm^{-2}}$. Our data,
hence, point against Stoner instability in our samples.

\begin{figure}[tbp]
\includegraphics[width=8cm]  {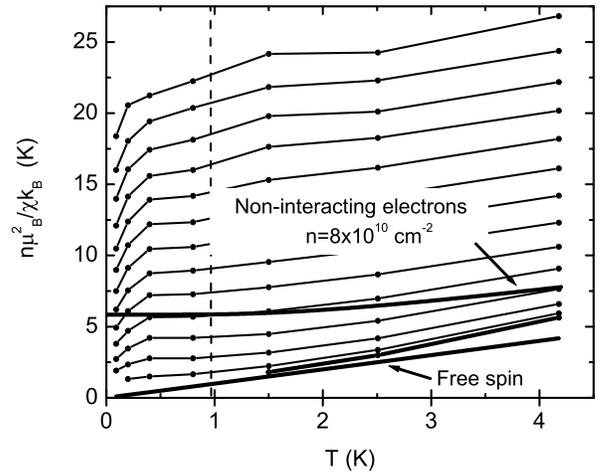}
\caption{Inverse susceptibility as determined from $M(B)$ at
$B=0.7$~T. Experimental points from bottom to top correspond to
densities $0.8\div 6\times 10^{11}$cm$^{-2}$ in $4\times
10^{10}~{\rm cm^{-2}}$ steps. The thick straight line depicts
Curie law and the dashed line marks $T=(g\mu_B/k_B)\times 0.7$~T.
The experimental points at $n=8\times 10^{10}$ are connected by a
thick line for comparison with the expectation for non-interacting
electrons of the same density.} \label{ki}
\end{figure}

Within effective medium theory the inverse susceptibility per
particle assumes Curie-Weiss form,
$\chi=\mu_{B}^{2}/k_{B}(T-T_{c})$. The value of $T_{c}$ in this
approximation provides an intuitive measure for the combined
effects of the kinetic energy and interaction. In particular, a
ferromagnetic instability requires positive $T_c$. The
paramagnetic nature of the 2D layer should, hence, be reflected in
the temperature dependence of the susceptibility $\chi$. The
inverse susceptibility, determined from $B=0.7$~T spin
magnetization, normalized by the electron density and expressed in
Kelvins is depicted as a function of temperature in Fig.~\ref{ki}.
For all densities the inverse susceptibility per spin is larger
(negative $T_{c}$) than the Curie value, $\chi ^{-1}=k_{B}T/\mu
_{B}^{2}$, indicating that in the balance between the Coulomb
energy gain and the kinetic energy toll associated with spin
polarization in the system, the latter wins. Yet, at the lowest
densities the victory is marginal, $T_{c}\cong0.2K$. To appreciate
the almost perfect balancing of the kinetic energy by the
interaction, we compare the data for $n=8\times 10^{10}~{\rm
cm}^{-2}$ (thick line connecting the data points in Fig.~\ref{ki})
to the theoretical inverse susceptibility of a non-interacting 2D
Fermi gas of the same density. The difference between the two
curves reflects the effects of the ferromagnetic interaction and
disorder, which are absent in an ideal non-interacting gas.
Remarkably, the susceptibility measured at densities just below
$n_{c}$, approaches the free spin one (Curie law) very closely,
implying that the kinetic energy is almost perfectly compensated
by the interaction. Yet, the former wins and paramagnetism
prevails. Since we believe that the free spin-like susceptibility
near the critical density is generic, rather than fortuitous, we
propose that the localization transition is driven, in addition to
disorder, by the strong exchange interaction, which promotes
localization through the Pauli principle. Localization, in turn,
reduces the overlap between the electron wave functions and,
hence, the exchange interaction (a strongly localized system is
believed to have a nearest neighbor  antiferromagnetic
order~\cite{Bhatt&Lee}). This scenario also explains the large
positive magnetoresistance observed in the vicinity of the
$n_c$~\cite{Pud parallel,Krav parallel}. Magnetic field aligns the
spins, and again by the Pauli principle, drives the system towards
the insulating phase. The localization transition at higher fields
is, hence, shifted to higher densities. The proposed scenario also
highlights the similarity between the localization transition in
high mobility 2DEG and the Mott transition~\cite{Mott}.

Note, that in contrast to all expectations, $\chi$ in the metallic
phase depends on temperature down to  0.1~K. This dependence
indicates the existence of a relevant energy scale considerably
smaller than the Zeeman one (dashed vertical line in
Fig.~\ref{ki}). Such an energy scale may originate from localized
spins which interact very weakly with each other. Quantification
of the number of localized spins and their contribution to
$M$~\cite{GD02} requires further study.

\begin{figure}[tbp]
\includegraphics [width=8cm]  {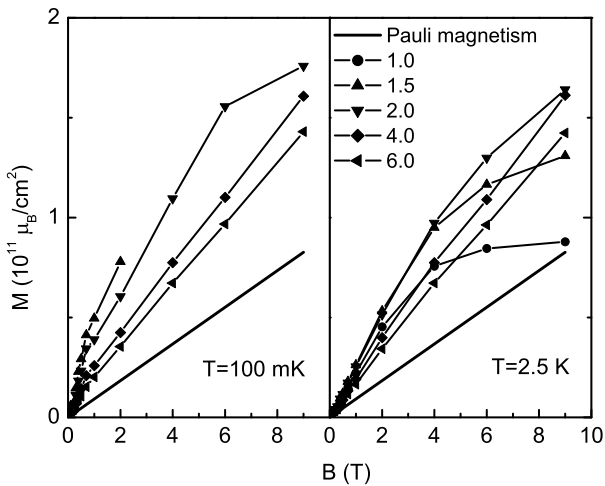}
\caption{Magnetization as a function of magnetic field at
$T=100$~mK and $T=2.5$~K. Densities are given in 10$^{11}~{\rm
cm^{-2}}$ units. Bold line - Pauli magnetization for
non-interacting degenerate fermions. As the temperature and
magnetic field are reduced, the magnetization becomes increasingly
nonlinear.} \label{M(B)}
\end{figure}

Fig.~\ref{M(B)} depicts the magnetization vs. magnetic field for
various densities. The magnetization at densities close to $n_c$
increases strongly with decreasing temperature. The magnetization
is nonlinear, implying that sufficiently low magnetic fields are
required (less than 0.7~T in our case) to determine the ``zero''
field susceptibility. At $T=100$~mK and $n=1.25 \times
10^{11}~{\rm cm}^{-2}$ the strong Coulomb interaction is
manifested in a 7.5 folds enhancement of the susceptibility
compared with non-interacting electrons. This susceptibility is
twice as large as the value extracted from the Sh-dH
data~\cite{gm} for the same density. We attribute the difference
to the localized states which persist into the metallic phase,
$n>n_c$, and are not sensed by the Sh-dH analysis. The weak
interaction between such spins should result in a very large
susceptibility at low temperatures. Indeed, as evident from
Figs.~\ref{ki} and \ref{M(B)}, the weak field susceptibility is
sensitive to temperature down to 100~mK. For stronger fields, for
which the Zeeman energy exceeds the temperature, almost all spins
are polarized and susceptibility depends very weakly on
temperature.

\section{Comparison with recent searches for the Stoner instability}

\begin{figure}[tbp]
\includegraphics [width=8cm] {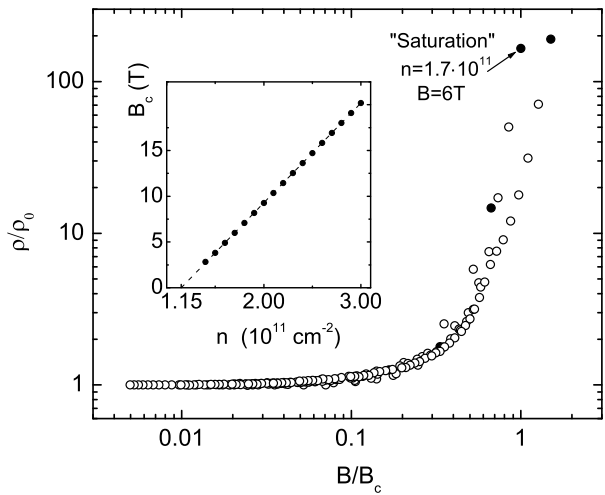}
\caption{Normalized magnetoresistance at different densities
plotted versus scaled magnetic field, $B/B_c(n)$. The field
$B_c(n)$, used to scale the data, is shown by dots in the inset.
The extrapolation of the scaling field to zero at a finite
density, $n_0$, was used in~\cite{SKDK} to claim a ferromagnetic
instability at $n=n_0$.} \label{pms}
\end{figure}

To the best of our knowledge, with the exception of
Refs.~\onlinecite{SKDK,VZ2}, there is no reported experimental
observation of Stoner instability in 2D systems. In particular,
recent susceptibility measurements based on Sh-dH
data~\cite{gm,Pudalov:02}, carried out down to the critical
density of a superb sample ($n_{c}=8\times10^{10}~{\rm cm^{-2}}$),
find a finite susceptibility in the whole density range, in
agreement with our result. We therefore turn to careful
examination of the arguments used in~\onlinecite{SKDK,VZ2} to
claim the observation of such an instability. Both references rely
on the magnetoresistance measured as a function of in-plane
magnetic field. At high densities the resistance grows
approximately quadratically with the field up to some density
dependent field. Then it saturates or at least becomes weakly
field dependent. It is believed, that at these densities the
saturation field corresponds to full spin polarization. This large
positive magnetoresistance is generic to all samples that show the
so-called metallic phase in 2D and is, hence, very likely to
provide an important clue for the understanding of the latter
phenomenon. The authors of~\onlinecite{SKDK,VZ2} have noticed that
normalized magnetoresistance curves, $\rho(n,B)/\rho(n,0)$
(magnetoconductance in the case of~\onlinecite{VZ2}), measured at
different densities, can be collapsed onto a single curve if the
field is scaled by a density dependent field $B_{c}(n)$ (we use
the notation of~\onlinecite{SKDK}. Ref.~\onlinecite{VZ2} utilizes
somewhat different analysis in the same spirit). Moreover, for
high densities, where magnetoresistance saturation is observed,
$B_{c}(n)$ can be set to the saturation field. At lower densities
(still above $n_{c}$) the saturation can no longer be observed but
a scaling field, $B_{c}(n)$, can still be found, so that the
curves collapse one on top of another. The authors of these
references noticed that $B_c$ vanishes approximately linearly when
the density approaches $n_{c}$, namely, $B_{c}(n)\propto n-n_{c}$.
They then argued that, since $B_{c}(n)$ corresponds to full spin
polarization at high densities, it should also correspond to full
polarization at lower densities, where magnetoresistance
saturation is no longer observable. Following that logic all the
way to the critical density, they concluded that the vanishing of
$B_{c}(n)$ at some finite density must indicate spontaneous
polarization at zero field, i.e. the long awaited Stoner
instability. We can not exclude Stoner instability in the superb
samples used in~\onlinecite{SKDK,VZ2}, but we can prove that the
procedure used to conclude the instability is wrong. To that end
we show in Fig.~\ref{pms} that our data obey the same scaling as
in~\onlinecite{SKDK}. In anticipation of the same dependence of
$B_{c}(n)$ upon density as in~\onlinecite{SKDK} we surmise
$B_{c}(n)\propto n-n_{0}$ (inset to Fig. 6) and find that for
$n_{0}=1.15\times10^{11}~{\rm cm^{-2}}$ all our scaled
magnetoresistance curves collapse onto a single curve
(Fig.~\ref{pms}). Following the arguments in~\onlinecite{SKDK,VZ2}
we could have concluded Stoner instability at $n_{0}$, but our
direct magnetization measurements at that density show finite
susceptibility. The same fact is also reflected in
Fig.~\ref{M(n)}b. If the high field magnetization is extrapolated
to zero field (dashed line) one may have erroneously predicted
instability. Carrying the measurements to smaller fields exclude
that possibility. The wrong assumption of
Refs.~\onlinecite{SKDK,VZ2} is the identification of $B_{c}(n)$
with a full polarization at all densities. Some of the authors
of~\onlinecite{VZ2} later restricted their conclusion to the
non-existing case of perfectly clean
samples~\cite{Vitkalov:prb02}.

In summary. Using a novel technique we were able to measure the
thermodynamic spin magnetization of strongly correlated 2D
electrons in a single 2D layer. Albeit the substantial enhancement
of the low temperature susceptibility, no ferromagnetic
instability was observed. Yet, at densities in the vicinity of the
critical one we observe almost free-spin like susceptibility,
indicating nearly perfect compensation of the kinetic energy by
the ferromagnetic interaction. The possible relation between the
large spin susceptibility at the critical density and the
transition to strong localization calls for further theoretical
and experimental studies. Understanding the role and nature of the
localized spins might also turn to be important.
\section{acknowledgements}
We have benefited greatly from valuable discussions with
A.~Finkelstein, A.~Stern, A.~Kamenev and V.~Dolgopolov. This work
was supported by the Israeli National Science Foundation, the DIP
foundation and INTAS. V.P. was supported by NATO, NSF, INTAS, and
Russian programs RFBR, ``Physics of nanostructures'',
``Integration of  education and academic research'', ``The State
support of leading scientific schools''.


\begin{thebibliography}{qqq}

\bibitem{Senatore}G.~Senatore et al., Sol. St. Commun. {\bf 119}, 333 (2001)

\bibitem{attaccalite}C.~Attaccalite, S.~Moroni, P.~Gori-Giorgi,
and G.~B.~Bachelet, Phys. Rev. Lett. {\bf 88}, 256601 (2002)

\bibitem{Don}See, e.g., S. Doniach and E.H. Sondheimer, {\em Green's
functions for Solid State Physicists} (W.A. Benjamin, Reading, MA,
1974)

\bibitem{T&C} B. Tanatar and D.M. Ceperley, Phys. Rev.
{\bf B} {\bf 39}, 5005 (1989)

\bibitem{Ceperley_2001}B.~Bernu, L.~Candido, D.~M.~Ceperley,
Phys. Rev. Lett., {\bf 86}, 870 (2001)

\bibitem{Voelker01} K.~Voelker, S.~Chakravarty, Phys. Rev. {\bf
B}, {\bf 64}, 235125 (2001)

\bibitem{gang4} E.~Abrahams, P.~W.~Anderson, D.~C.~Licciardello
 and T.~V.~Ramakrishnan, Phys. Rev. Lett., {\bf 42} 673 (1979)

\bibitem{aa}B.~L.~Altshuler and A.~G.~Aronov,  in
{\it Electron-electron interactions in disordered systems}, ed. by
A.~ L.~Efros and M.~Pollak  (Elsevier, Amsterdam, 1985), p. 1

\bibitem{Fin} A.~M.~Finkelstein, Sov.~Phys.~JETP {\bf 57}, 97 (1983)

\bibitem{Zala} G\`{a}bor Zala, B.N. Narozhny, and I.L. Aleiner, Phys. Rev. {\bf B} 64,
201201 (2001).

\bibitem{Punnoose} A.~Punnoose and A.~M.~Finkelstein, {\it Phys. Rev.
Lett.} 88, 016802 (2001); Physica A 302, 318 (2001)

\bibitem{FS68} F.~F.~Fang and P.~J.~Stiles, Phys. Rev. {\bf 174},
823 (1968).
\bibitem{OH99} T.\ Okamoto, K.\ Hosoya, S.\ Kawaji, and A.\ Yagi,
Phys. Rev. Lett. {\bf 82}, 3875 (1999).

\bibitem{VZ1} S.~A.~Vitkalov, H. Zheng, K.~M.~Mertes, M.~P.~Sarachik, T.~M.~Klapwijk,
Phys. Rev. Lett. {\bf 85}, 2164 (2000).

\bibitem{gm}V.~M.~Pudalov, M.~E.~Gershenson, H.~Kojima, N.~Butch,
E.~M.~Dizhur, G.~Brunthaler, A.~Prinz, G.~Bauer, Phys. Rev. Lett.
{\bf 88}, 196404 (2002), cond-mat/0105081.

\bibitem{SKDK} A.~A.~Shashkin, S.~V.~Kravchenko, V.~T.~Dolgopolov, T.~M.~Klapwijk,  Phys. Rev. Lett., {\bf 87}, 86801
(2001)

\bibitem{VZ2} S.~A.~Vitkalov, H. Zheng, K.~M.~Mertes, M.~P.~Sarachik, T.~M.~Klapwijk,
Phys. Rev. Lett. {\bf 87}, 086401 (2001).

\bibitem{Pudalov:02}V.~Pudalov, M.~E.~Gershenson, and H. Kojima,
cond-mat/0110160.

\bibitem{Meinel:APL97} I.~Meinel, D~Grundler, S.~Bargst\"{a}dt-Franke, C.~Heyn,
D.~Heitmann, and B.~David, Appl. Phys. Lett. {\bf 70}, 3305
(1997).

\bibitem{Awshalom00} J.~G.~E.~Harris D.~D.~Awschalom, K.~D.~Maranowski and
A.~C.~Gossard,  Journ. Appl. Phys., {\bf 87}, 5102 (2000).

\bibitem{rhoT1} S. V. Kravchenko, G. V. Kravchenko, J. E. Furneaux, V. M. Pudalov,
M. D'Iorio, {\it Phys. Rev. B} {\bf 50}, 8039 (1994)
\bibitem{rhoT2} S.\ V.\ Kravchenko {\it et al.}, Phys. Rev. {\bf B} {\bf 51},
7038 (1995)
\bibitem{Prus02} O.~Prus, M.~Reznikov, U.~Sivan, V.~Pudalov, Phys. Rev. Lett. {\bf 88}, 16801 (2002)

\bibitem{AbrKravSar}E. Abrahams, S. V. Kravchenko, M. P. Sarachik,
Rev. Mod. Phys. {\bf 73}, 251 (2001)

\bibitem{Bhatt&Lee} R.~N.~Bhatt and P.~A.~Lee, Phys. Rev. Lett. {\bf 48}, 344
(1982)

\bibitem{Pud parallel} V.~M.~Pudalov, G. Brunthaler, A. Prinz, and G. Bauer,
JETP Lett.\ {\bf 65}, 932 (1997).

\bibitem{Krav parallel} D.~Simonian, S.~V. Kravchenko, M.~P. Sarachik,
 and V.~M. Pudalov, Phys.~Rev.~Lett.~{\bf 79}, 2304 (1997).

\bibitem{Mott} N.~F.~Mott, {\em Metal-Insulator Transitions},
(Taylor \& Francis, London, 1990)
\bibitem{GD02} A. Gold and V.~T.~Dolgopolov, J. Phys.: Condens. Matter {\bf 14}, 7091 (2002)

\bibitem{Vitkalov:prb02} S.~A.~Vitkalov, M.~P.~Sarachik and T. M.
Klapwijk, Phys. Rev. B {\bf 65}, 201106 (2002)

\end{thebibliography}
\end{document}